\documentclass[sn-mathphys,Numbered]{sn-jnl}
\usepackage{graphicx}%
\usepackage{multirow}%
\usepackage{amsmath,amssymb,amsfonts}%
\usepackage{amsthm}%
\usepackage{mathrsfs}%
\usepackage[title]{appendix}%
\usepackage{xcolor}%
\usepackage{textcomp}%
\usepackage{manyfoot}%
\usepackage{booktabs}%
\usepackage{algorithm}%
\usepackage{algorithmicx}%
\usepackage{algpseudocode}%
\usepackage{listings}%
\begin{document}

\title[Article Title]{Segmentation of tibiofemoral joint tissues from knee MRI using MtRA-Unet and incorporating shape information: Data from the Osteoarthritis Initiative}

\author*[1]{\fnm{Akshay} \sur{Daydar}}\email{adaydar@iitg.ac.in}

\author[2]{\fnm{Alik} \sur{Pramanick}}\email{p.alik@iitg.ac.in}

\author[2]{\fnm{Arijit} \sur{Sur}}\email{arijit@iitg.ac.in}

\author[1]{\fnm{Subramani} \sur{Kanagaraj}}\email{kanagaraj@iitg.ac.in}

\affil[1]{\orgdiv{Department of Mechanical Engineering}, \orgname{Indian Institute of Technology Guwahati}, \orgaddress{\city{Kamrup}, \postcode{781039}, \state{Assam}, \country{India}}}

\affil[2]{\orgdiv{Department of Computer Science and Engineering}, \orgname{Indian Institute of Technology Guwahati}, \orgaddress{\city{Kamrup}, \postcode{781039}, \state{Assam}, \country{India}}}

\abstract{Knee Osteoarthritis (KOA) is the third most prevalent Musculoskeletal Disorder (MSD) after neck and back pain. To monitor such a severe MSD, a segmentation map of the femur, tibia and tibiofemoral cartilage is usually accessed using the automated segmentation algorithm from the Magnetic Resonance Imaging (MRI) of the knee. But, in recent works, such segmentation is conceivable only from the multistage framework thus creating data handling issues and needing continuous manual inference rendering it unable to make a quick and precise clinical diagnosis. In order to solve these issues, in this paper the Multi-Resolution Attentive-Unet (MtRA-Unet) is proposed to segment the femur, tibia and tibiofemoral cartilage automatically. The proposed work has included a novel Multi-Resolution Feature Fusion (MRFF) and Shape Reconstruction (SR) loss that focuses on multi-contextual information and structural anatomical details of the femur, tibia and tibiofemoral cartilage. Unlike previous approaches, the proposed work is a single-stage and end-to-end framework producing a Dice Similarity Coefficient (DSC) of 98.5\% for the femur, 98.4\% for the tibia, 89.1\% for Femoral Cartilage (FC) and 86.1\% for Tibial Cartilage (TC) for critical MRI slices that can be helpful to clinicians for KOA grading. The time to segment MRI volume (160 slices) per subject is 22 sec. which is one of the fastest among state-of-the-art.
Moreover, comprehensive experimentation on the segmentation of FC and TC which is of utmost importance for morphology-based studies to check KOA progression reveals that the proposed method has produced an excellent result with binary segmentation.} 


\maketitle
\section{Introduction}\label{sec1}
Knee Osteoarthritis (KOA) is one of the most prevalent Musculoskeletal Disorders (MSD) affecting the lower limb. According to the global burden of disease (2019) data, about 1.7 billion people are affected by a work-related MSD, with 31\% (528 million) of them suffering from osteoarthritis \cite{cieza2020global}. Depending on the size of the older population, the prevalence of osteoarthritis is expected to climb as average life expectancy rises. KOA is a disorder with medical and psychological effects on individuals, as well as societal and economic implications \cite{litwic2013epidemiology}. The diagnosis and severity assessment of KOA in clinical research is primarily based on radiology data. Due to the inherent limitation of the X-rays in determining the structural course of KOA, MRI is clinically preferred that is used to obtain 2D and 3D images of intra-articular soft-tissue structures, including tibiofemoral cartilage and bones to detect structural abnormalities in the knee \cite{deng2021coarse}.

Segmentation is a common tool to detect structural abnormalities in the knee joint from MRI. Specifically, cartilage segmentation is used to compute the rate of cartilage degradation and calculate the morphology parameters while bone segmentation is used to obtain bone models for analyzing the bio-mechanical stresses at varied knee sites and keep track of changes in bone shape and structure brought on by structural deformation \cite{ahmed2022comprehensive}. Although manual knee joint segmentation is considered the gold standard, it presents several challenges because of (1) the variability of pathological structures regarding their shape, size, and spatial resolution, and (2) the uncertainties involved in accurately defining both inter- and intra-cartilage boundaries, which can involve areas where cartilage surfaces touch, intensity in-homogeneity, and cartilage loss resulting from the progression of the disease. Hence manual segmentation of MR images is often regarded as a time-consuming and labor-intensive task that needs clinicians' knowledge and experience which is highly subjective. To overcome this problem, several Convolutional Neural Networks (CNN) approaches \cite{ambellan2019automated},\cite{deng2021coarse} have been recently developed to automate the segmentation process. A CNN is a deep neural network that is trained on given data while minimizing specific loss functions such as pixel loss (e.g. Dice loss \cite{sorensen1948method}), cross-entropy loss \cite{shannon1948mathematical}. Recently, some research studies have explored the use of perceptual loss in medical imaging \cite{al2017shape},\cite{mosinska2018beyond},\cite{kim2019mumford},\cite{lambert2021geometrically} to address segmentation consistency issues. Perceptual loss is employed to apply transformations to the ground truth, aiming to derive representations that effectively highlight the geometric and contour characteristics of the object of interest. Perceptual loss can be perceived as the regularization term added to the existing loss function \cite{sarasaen2021fine} which can be any of the following forms: Shape, topology, size, inter-region priors. \cite{el2021high}.

The proposed work is a discriminative network that learns the multilevel context using a {Multi-resolution Feature Fusion (MRFF)} module while focusing on small tissue details using the Convolutional Block Attention Module (CBAM) while being trained on the combined loss function (pixel-wise and shape reconstruction loss) to rectify semantic and structural errors respectively. The paper is organized into five sections, the first section highlights an introduction and major literature related to MRI segmentation in KOA, the second section describes the material and methods, and the third section focuses on experimental qualitative and qualitative results then followed by an ablation study and discussion section.

\section{Related Works}\label{sec2}

In order to segment multiple knee joint tissues, various semi-automated and automated segmentation systems have been proposed with varying degrees of complexity and accuracy. Semiautomatic methods like those using active shape models \cite{lynch2001automating}, \cite{fripp2009automatic} and active appearance models \cite{williams2010automatic} utilize anatomical insights through geometric priors, enabling reliable segmentation, even when dealing with artifacts or images with low contrast. Such segmentation algorithm relies on the statistical shape model (SSM) to limit segmentation to plausible shapes that result in rigorous registration and feature matching stages, and greater processing time, making it inappropriate for the non-linear shape changes segmentation task.

Today, CNN is one of the most promising automated segmentation techniques. Due to outstanding automatic feature learning capacity, CNN-based techniques can easily handle non-linear shape changes thus overcoming rigorous feature matching and registration steps. Current state-of-the-art CNN-based methods concentrate on segmenting slice-wise two-dimensional (2D CNN) approaches \cite{liu2018deep}, three-dimensional/volumetric methods (3D CNN) \cite{ambellan2019automated}, and training 2D and 3D CNNs with a semi-supervised learning framework \cite{burton2020semi}. 
Ambellan et al.\cite{ambellan2019automated} combined the statistical shape modelling with the CNN on the 507 manually segmented femur, tibia and tibiofemoral cartilages (also known as the OAI-ZIB dataset). The regularization of shape prior and 2D-3D context learning was achieved using SSM and CNN respectively. Additionally, SSM is employed to remove segmentation bias by removing false positives in sub-volume segmentation of the femur and tibia. 
Later on, the OAI-ZIB dataset was considered a benchmark for comparing performance among different studies. In addition, SK10 \cite{heimann2010segmentation} and Imorphics were used for training and validation of segmentation due to their openly accessible status. In a few research works \cite{kessler2020optimisation},\cite{li2022entropy}, in-house datasets were also used for verification purposes. Kessler et al.\cite{kessler2020optimisation} studied the segmentation of knee tissues using conditional Generative Adversarial Networks (cGANs) with different objective functions and discriminator receptive field sizes. Latif et al.\cite{abd2021automated} provided a completely automated segmentation pipeline based on a 2D-3D ensemble {Unet} model with foreground class oversampling, deep supervision loss branches, and gaussian weighted softmax score aggregation, among other features. Li et al. \cite{li2022entropy} proposed a three-stage segmentation pipeline, that included entropy and distance maps to refine the segmentation results of nn-Unet model. Deng et al. \cite{deng2021coarse} suggested a two-stage "coarse-to-fine segmentation" network employing Unet++ \cite{zhou2019unet++} twice for each tissue to locate an accurate Region of Interest (ROI) bounding boxes and segmentation refinement respectively. In some of the recent works related to cartilage segmentation, Ebrahimkhani et al. \cite{ebrahimkhani2022automated} developed a Joint diffeomorphic hand-crafted learning (JD-HCl) based pipeline that incorporates dual stage; 3D Unet-based CNN for 3D context learning, diffeomorphic mapping for spatial image alignment, and a hand-crafted feature-based multi-class (RUSBoost) classifier model and Liang et al. \cite{liang2022position} implemented the Position prior Clustering based self-Attention Module (PCAM) to capture long-distance relationships between class centers and feature locations, in an attempt to address the limited receptive field size.

Overall, in most of the works, the two-stage segmentation pipeline is adopted. The first stage of such a pipeline attempts to capture spatial details, possibly in 2D and 3D perspective while the second stage is utilized to refine the segmentation results of the first stage. However, tackling the segmentation problem from a 2D perspective has achieved success over the past few years and produced excellent results with the use of the attention schemes on simple baseline architectures but handling a 3D context is found to be computationally expensive over limited performance improvement \cite{abd2021automated}. Furthermore, due to offline implementations, data-dependent results and post-processing phases including morphological operations, eliminating small islands or unconnected regions, and Conditional Random Fields (CRF) are observed to be time-consuming in most implementations. In addition, in the second stage the incorporation of shape information using diffeomorphic mapping \cite{ebrahimkhani2022automated}, entropy and distance maps \cite{li2022entropy} is found to be inaccurate at the boundary interface. It hence acts as a bottleneck for the segmentation process. Accordingly, the proposed method is intended to overcome segmentation difficulties, notably, those caused by inconsistencies in the capturing context of multiple tissues, irregular geometries of smaller tissue structures, and uncertainties in anatomical shape and delineating inter-and intra-cartilage boundaries.

In order to solve the above issues, the main contributions of the proposed work are as follows;
\begin{enumerate}
    \item A Multi-resolution Feature Fusion (MRFF) block is proposed that is accompanied by CBAM 
    focuses on multilevel spatial and channel contexts for accounting relevant local and global information for accurate segmentation of femur, tibia and tibiofemoral cartilage. It is found that aggregation of these modules improves the segmentation performance in terms of overall segmentation evaluation metrics.
    \item  A Shape Reconstruction (SR) loss is proposed to take into account structural information. It is found that aggregation of proposed loss with pixel-wise loss function helped specifically in the case of knee MR images to restore fine anatomical details of the femur and tibia.
    \item The proposed Multi-resolution Attentive-Unet (MtRA-Unet) model is validated for binary segmentation of cartilages. It is found that the MtRA-Unet model with the combined loss function achieved better performance overall.
    
\end{enumerate}
Extensive experiments are performed for the varied loss functions, weighted schemes for loss functions, and different MRFF modules to verify the robustness of the proposed work.

\section{Materials and Methods}
\subsection{Data Collection}
The input MR images and segmented masks were collected from Osteoarthritis Initiative (OAI) \cite{nih} and OAI-ZIB \cite{zibPublicData} repositories respectively. The OAI dataset contains knee MRIs from 4796 subjects that were acquired on Siemens 3 T Trio systems using a 3D double echo steady-state (DESS) sequence with water excitation. In the OAI dataset, each MRI volume includes 160 slices (images) in total.
The OAI-ZIB was derived from the OAI dataset, manually annotated at Zuse Institute Berlin (ZIB) \cite{ambellan2019automated}. It consists of 507 segmented volumes from MRI for Femoral Cartilage (FC), Tibial Cartilage (TC), Femoral Bone (FB) and Tibial Bone (TB) as reported in Table \ref{tab:T12}. The OAI-ZIB data covered all degrees of KOA (KL 0–4), with severe KOA (KL 3) being more common. Each voxel is coded as one of the four anatomical structures (FB: 1, FC: 2, TB: 3, and TC: 4), and the background code is set as 0.
\begin{table}
\caption{Summary of the employed OAI-ZIB Dataset \cite{ambellan2019automated}}
\label{tab:T12}
\centering
\begin{tabular}{lc}
\toprule
 \textbf{Parameter} & \textbf{Value}\\
\midrule
MRI scanner & Siemens 3T Trio \\
Acquisition & Sagittal \\
Image resolution (mm) & 0.36*0.36*0.7 \\
Number of subjects & 507 \\
Volume size (slices*pixels) & 160*(384*384)  \\
\botrule
\end{tabular}
\end{table}

\subsection{Methods}
The proposed scheme comprises the MtRA-Unet architecture and a shape reconstruction loss. Initially, the MtRA-Unet architecture undergoes training in an end-to-end fashion, utilizing a combined pixel-wise and shape reconstruction loss. Subsequently, the architecture is tested exclusively on slices deemed critical to KOA based on the criterion established by Deng et al. \cite{deng2021coarse}.\\

\subsection{Proposed MtRA-Unet Architecture}
\begin{figure*}
    \centering
    \includegraphics[scale=0.8]{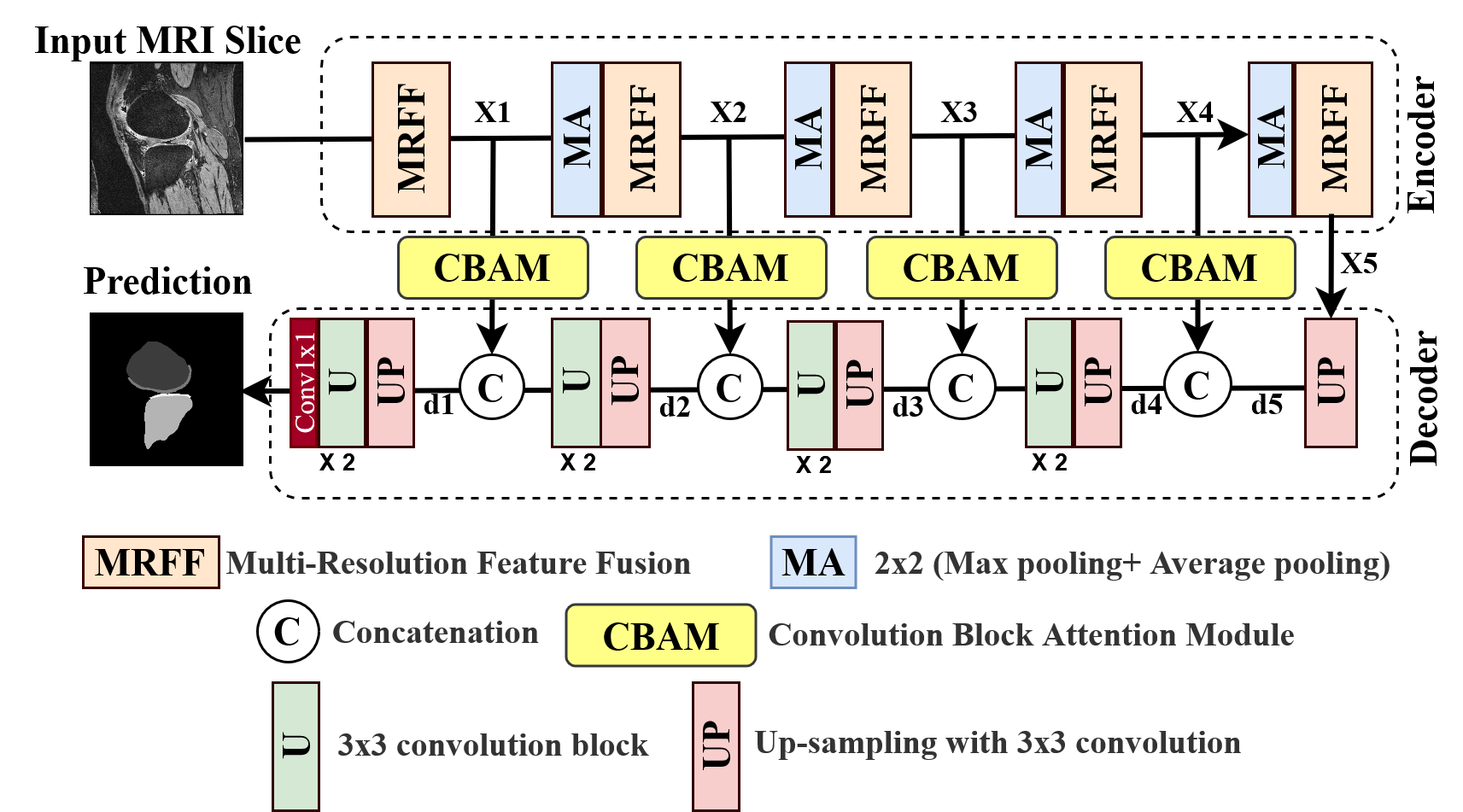}
    \caption{Illustration of the proposed Multi-Resolution Attentive-Unet (MtRA-Unet) architecture; Multi-Resolution Feature Fusion (MRFF) (see Figure \ref{fig:MRFF} module is incorporated in place of convolutions of Unet-encoder and Convolution Block Attention Module (CBAM) (see Figure \ref{fig:CBAM} is attached at the skip connections}
    \label{fig:Proposed Architecture}
\end{figure*}

The Unet \cite{ronneberger2015u} is selected as the baseline for constructing the MtRA-Unet model. The Unet \cite{ronneberger2015u} was proposed for the segmentation of microscopy cell images. Later, it found most of its applications in biomedical computer vision tasks. The {Unet} is an encoder-decoder architecture
containing skip connections, which can merge low-level and high-level features. However, local
attention may be a concern with these encoder-decoder designs for accurate image segmentation.
The feature-map grid in UNet is gradually down-sampled to capture a sufficiently large receptive
field and thus only the global context of location and relationship between tissues are considered.
However, the challenge of mitigating false-positive predictions for small objects exhibiting significant shape variability persists. In order to correctly locate the small tissue structure for accurate segmentation and fine-tune the global and local information, the "MtRA-Unet" model is proposed. The MtRA-Unet model is essentially the same as the Unet architecture with 5 encoding and decoding layers with some qualitative improvements, such as passing the skip connection through the CBAM and converting max pooling to a combination of max pooling and average pooling. Such a modified framework is named $\text{baseline}^{ \dagger}$. The $\text{baseline}^{ \dagger}$ is paired with the proposed MRFF module named as the "MtRA-Unet" model as shown in figure \ref{fig:Proposed Architecture}.\\ 

\begin{figure*}
    \centering
    \includegraphics[scale=0.7]{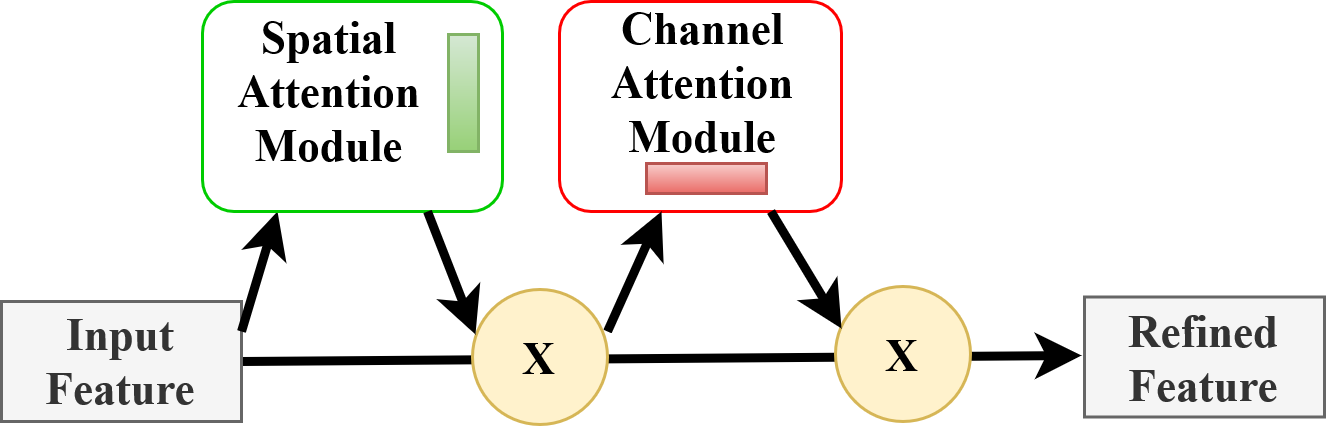}
    \caption{Schematic of the Convolutional Block Attention Module (CBAM) utilized in the paper}
    \label{fig:CBAM}
\end{figure*}

\noindent{\textit{CBAM}}\\
To accurately segment the smaller tissues such as cartilages the skip connections are passed through the CBAM \cite{woo2018cbam}. The CBAM \cite{woo2018cbam} sequentially 
intervenes the attention maps in channel and spatial dimensions, then the attention maps are multiplied to the input feature map for adaptive feature refinement as represented in Equation \ref{eqn:CBAMC} and \ref{eqn:CBAMS} respectively.
It may be noted that the CBAM in this paper consists of spatial attention followed by channel attention in series as shown in Figure \ref{fig:CBAM}.

\begin{equation}
\begin{aligned}
\label{eqn:CBAMC}
F_1 = M_c(F) \times F
\end{aligned}
\end{equation}

Where $F$, $F_1$ and $F_2$ are the input, intermediate and output feature maps respectively. $M_c$ and $M_s$ are the channel and spatial attention modules respectively. The $\times$ denotes channel-wise multiplication.
\begin{equation}
\begin{aligned}
\label{eqn:CBAMS}
F_2 = M_c(F_1) \times F_1 
\end{aligned}
\end{equation}

\noindent{\textit{GAP}}\\
{To assign channel-wise importance from the input feature maps, the Global Average Pooling (GAP) module is utilized in the MRFF. The GAP module computes the average of each input channels thus reducing the spatial dimensions of input while retaining their important information, ultimately transforming these feature maps into one-dimensional vector as indicated in Equation} \ref{eqn:GAP}.

\begin{equation}
\begin{aligned}
\label{eqn:GAP}
{GAP}(F_j) = \frac{1}{h \times w} \sum_{x=1}^{h} \sum_{y=1}^{w} F_j[h, w]
\end{aligned}
\end{equation}

\noindent{{Where, $GAP(F_j)$ is the output value for the $j^{th}$ feature map after Global Average Pooling and $F_j[h,w]$ represents the value of the $j^{th}$ feature map at spatial location (x,y).}}\\

\begin{figure*}
    \centering
    \includegraphics[scale=0.8]{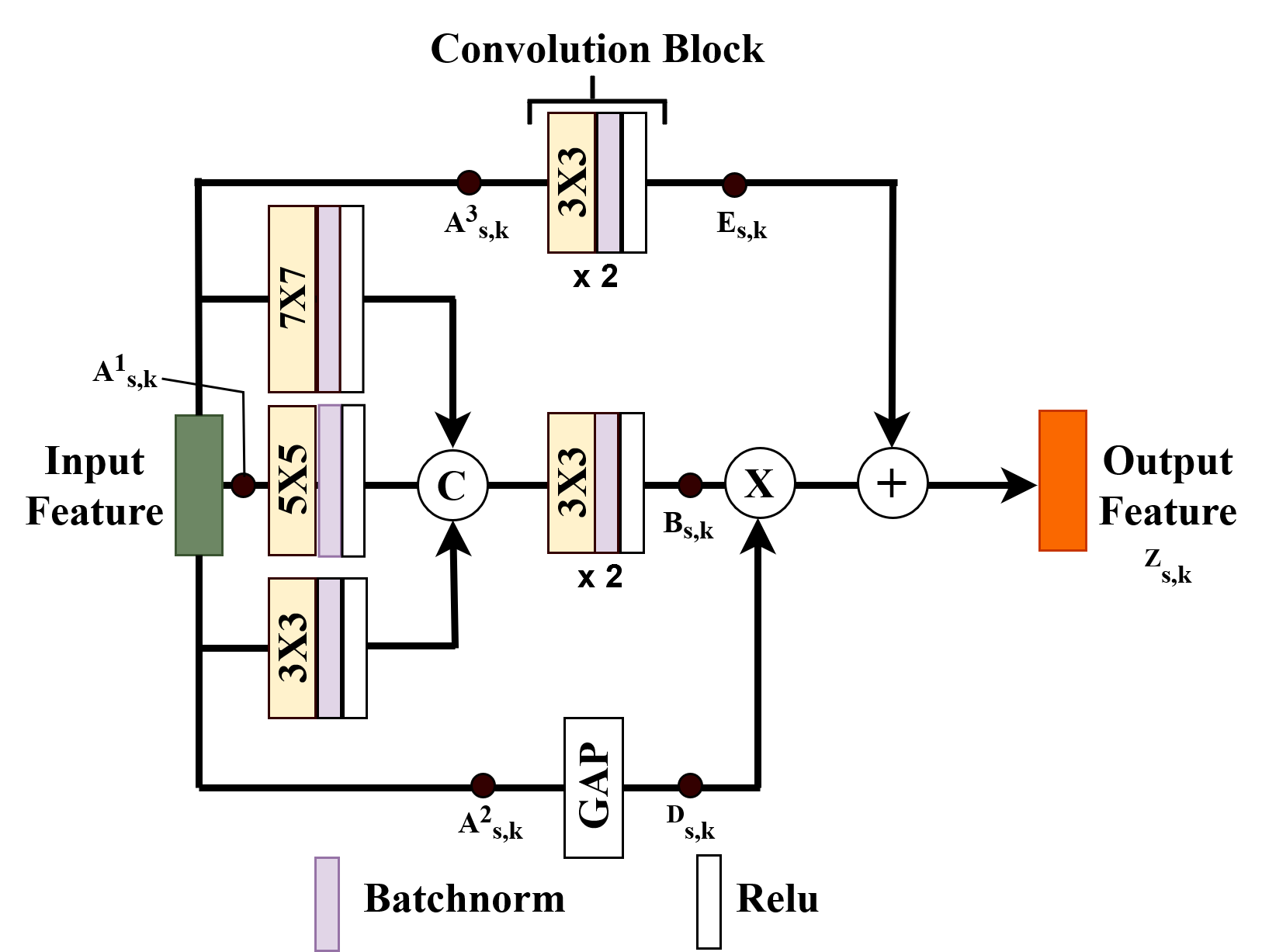}
    \caption{Schematic of the proposed Multi-Resolution Feature Fusion (MRFF) Module}
    \label{fig:MRFF}
\end{figure*}

\noindent{\textit{Proposed MRFF}}\\
To accurately contemplate global and local attention, the convolution block of the encoder is replaced with the proposed MRFF module. A convolution block consists of a convolution operation followed by batchnorm and Relu with a single repetition in the same order. The proposed MRFF module includes varied feature maps to take into account multi-scale contexts that are connected in parallel with the GAP module. Given $s^{th}$ slice of the $k^{th}$ subject, the input feature map $ A^1_{s,k}$ with the dimension $C*H*W$ (where $C$ are a number of channels, $H$ and $W$ are dimensions of the feature map) is passed through multiple feature kernels such as $3*3$, $5*5$ and $7*7$. The concatenated feature map ($B_{s,k}$) is then passed through the $3*3$ convolution block as shown in Figure \ref{fig:MRFF}.  The mathematical expression is represented in Equation \ref{eqn:equMRFF1}.

\begin{equation}
\begin{aligned}
\label{eqn:equMRFF1}
B_{s,k} = Relu(w^p[concat(T^1_{s,k},T^2_{s,k},T^3_{s,k}] + b^p)
\end{aligned}
\end{equation}
Where, $T^1_{s,k},T^2_{s,k},T^3_{s,k}$ are feature maps obtained after $3*3$,$5*5$ and $7*7$ convolution operation.\\
The first copy of the feature map $A^2_{s,k}$, is passed through the GAP module while the second copy $A^3_{s,k}$ is passed through $3*3$ convolution blocks that helped in enriching the feature content. The feature maps obtained from the above operations ($D_{s,k}$ and $E_{s,k}$) are framed into a mathematical setup to derive the final feature map ($Z_{s,k}$) from MRFF module as represented in Equation \ref{eqn:equMRFF}.

\begin{equation}
\begin{aligned}
\label{eqn:equMRFF}
Z_{s,k} = (B_{s,k} * D_{s,k}+ E_{s,k})
\end{aligned}
\end{equation}

\subsection{Loss Functions}
\noindent\textit{Pixel-wise loss functions:}
To consider the semantic inaccuracies, the weighted dice loss \cite{sorensen1948method} with weighted cross entropy loss \cite{shannon1948mathematical} is utilized as represented in Equations \ref{equ:equpx1} and \ref{equ:equpx2};
\begin{equation}
\begin{aligned}
\label{equ:equpx1}
L_{WCL,j} = \sum_{j=1}^{m}\alpha_{j}[(-Y_{j}\log(K_{j}+
(1 - \frac{K_{j}\cap Y_{j}} {\lvert K_{j} \rvert  +\lvert Y_{j} \rvert})], j=1,2,...,m
\end{aligned}
\end{equation}

$L_{WCL,j}$ and $\alpha_{i}$ denote the weighted pixel-wise loss and channel weight for $j^{th}$ channel.

\begin{equation}
\begin{aligned}
\label{equ:equpx2}
L_{WCL} = \alpha \sum_{j=1}^{m}(-Y_{j}\log(K_{j}+
(1 - \frac{K_{j}\cap Y_{j}} {\lvert K_{j} \rvert  +\lvert Y_{j} \rvert})
\end{aligned}
\end{equation}

Where $L_{WCL}$ is the weighted cross entropy \cite{shannon1948mathematical} and weighted dice loss \cite{sorensen1948method} for m channels and $\alpha$ denotes the channel weight vector of dimension 1xm.\\

\begin{figure*}
    \centering
    \includegraphics[scale=0.65]{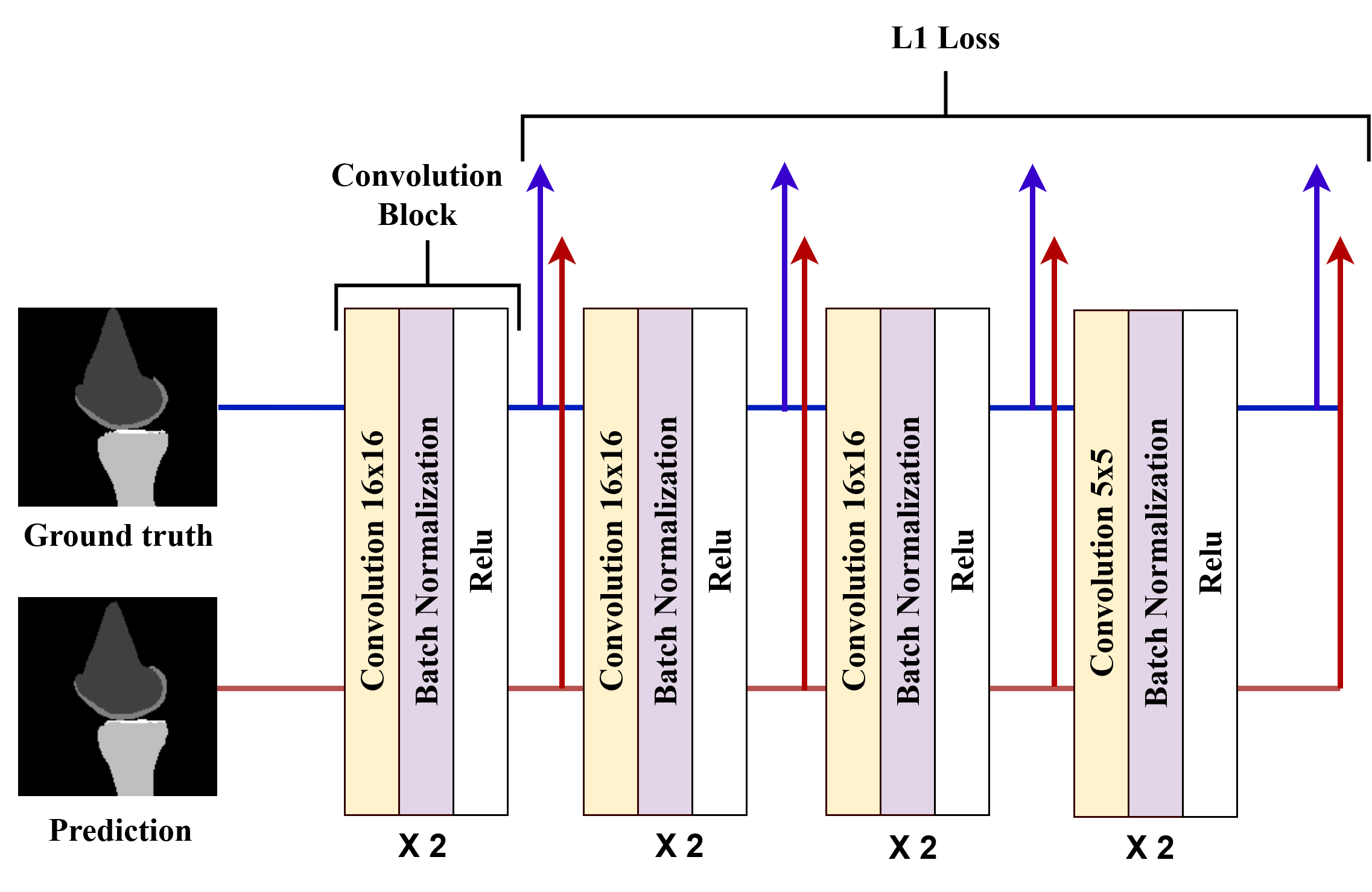}
    \caption{Illustration of the proposed Shape Reconstruction (SR) loss function}
    \label{fig:loss}
\end{figure*}

\textit{Proposed Shape reconstruction loss Function:}\\
\setlength{\parindent}{10pt} To consider the structural information along with semantic details from the predicted mask, a novel SR loss function is proposed. The proposed loss function is inspired by the part of CycleNet \cite{zhang2020improved} that calculates the feature difference between the predicted mask and ground truth at varied network depths. The L1 Loss is used for the calculation of feature difference as shown in figure \ref{fig:loss}. The proposed loss function consists of 4 convolution operations with three 16*16 convolution blocks in the beginning and middle and one 5*5 convolution block at the end. The mathematical formulation of the proposed loss is represented in Equations \ref{equ:equSRl1} and \ref{equ:equSRL}.

\begin{equation}
\begin{aligned}
\label{equ:equSRl1}
L_{SRL,i} = \sum_{i=1}^{n}\lambda_{i}(\lvert K_{i} -Y_{i} \rvert), i=1,2,...,n
\end{aligned}
\end{equation}

Where K and Y are predictions and ground truth respectively. $L_{SRL,i}$ and $\lambda_{i}$ denote the shape reconstruction loss and layer weight for $i^{th}$ layer.

\begin{equation}
\begin{aligned}
\label{equ:equSRL}
L_{SRL} = \lambda\sum_{i=1}^{n}(\lvert K_{i} -Y_{i} \rvert)
\end{aligned}
\end{equation}

Where $L_{SRL}$ is the shape reconstruction loss for n layers and $\lambda$ is the weight vector of dimension 1xn.
The combination of shape reconstruction and pixel-wise loss function is accounted to train the "MtRA-Unet" model. The mathematical formulation of the combined loss function is reported in Equation \ref{equ:equcom} as follows;

\begin{equation}
\begin{aligned}
\label{equ:equcom}
L_{T} = \gamma L_{WCL} + \eta L_{SRL}
\end{aligned}
\end{equation}

Where $\gamma$ and $\eta$ are the weights for pixel and shape reconstruction loss functions.

\subsection{ROI computation and strategy to reduce false positives}
\begin{figure*}
    \centering
    \includegraphics[scale=0.4]{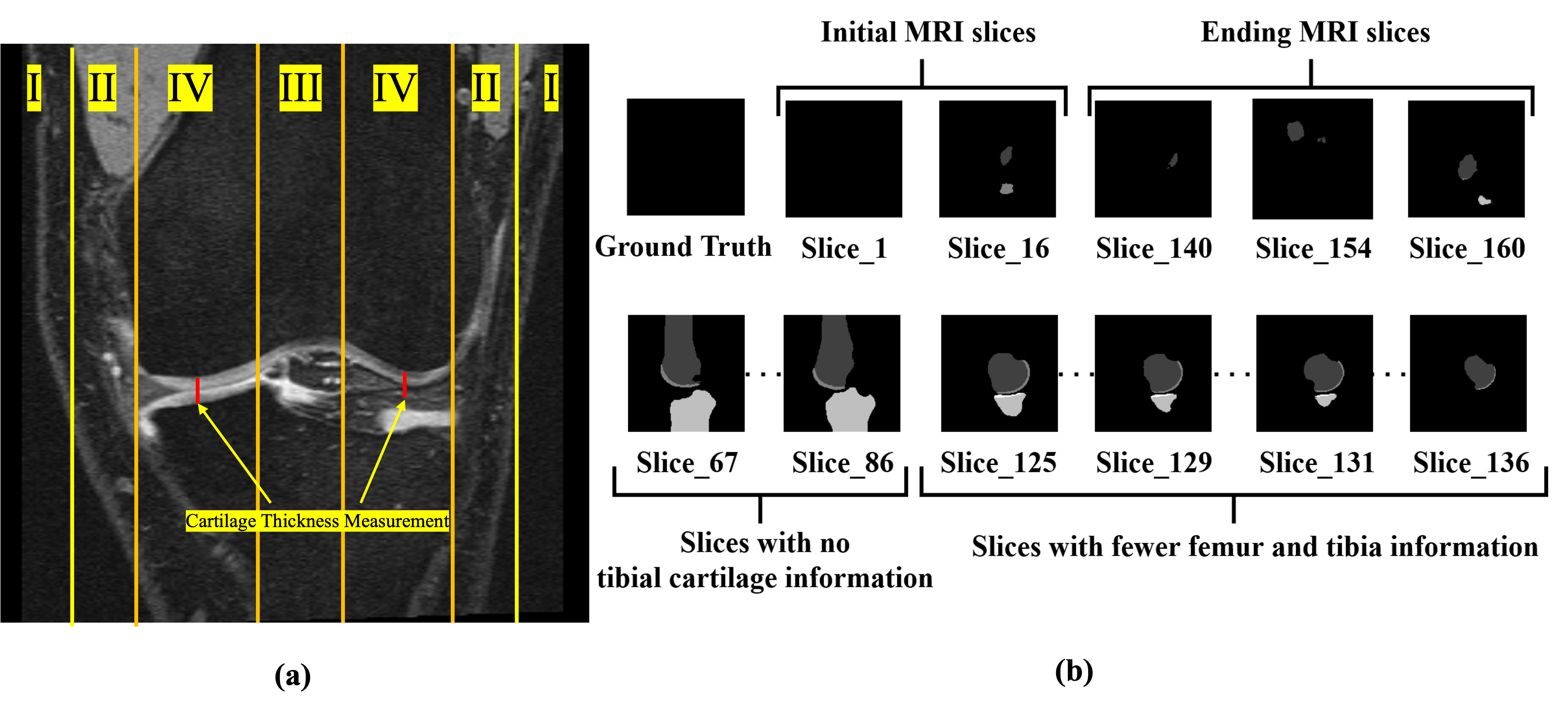}
    \caption{ (a) Coronal and (b) Sagittal view of MR slices with the depiction of (I) initial and ending slices (i.e. noise-only slices with no tissue information), (II) slices with fewer occupancy of femoral and tibial bone, and (III) slices with anterior and posterior cruciate ligaments (that contain no tibia cartilage information), and (IV) medial-lateral regions (selected for segmentation analysis) for subject ID: 9471287} 
    \label{fig:False Positives}
\end{figure*}

To improve diagnosis efficacy and lessen false positives for prediction in KOA, it is required to concentrate more on the ROI slices that contain femur, tibia, and tibiofemoral cartilage. To fulfill these requirements, the criterion of Deng et al.\cite{deng2021coarse} was found to be appropriate. According to the criterion, the performance evaluation must be carried out on only those slices that meet the minimum pixel count rule. The threshold in this rule is chosen in such a way that it can confine those ROI slices that contain only the weight-bearing region in the knee. The weight-bearing region is identified as 20\% of the total maximum medial-lateral cartilage width for each condyle \cite{koo2005considerations}, \cite{andriacchi2009gait} during walking, as illustrated in Figure \ref{fig:False Positives} (A), where the cartilage thickness measurement has significant influence in KOA grading \cite{koo2005considerations}. This criterion is observed to filter out (1) the slices with no tissue information (initial and ending MRI slices) that often produce false positive results, (2) the slices with fewer occupancy of femoral and tibial bone pixels and (3) the slices with anterior and posterior cruciate ligaments that have no tibial cartilage information when the image size is adjusted to 150*150 as represented in Figure \ref{fig:False Positives} (B). It is important to note that, the segmentation from the “MtRA-Unet” Model is performed for all the knee MRI slices per volume. However, the evaluation is measured only for those slices that are critical in the diagnosis of KOA. 

\subsection{Evaluation Metrics}
\noindent{\textit{Quantitative metrics}}\\
To measure the segmentation accuracy Dice Similarity Coefficient (DSC), Volumetric Error (VOE), Hausdorff Distance (HD), and Pixel Accuracy (PA) for each sample are studied. The DSC (\%) is the most commonly used evaluation metric for medical image segmentation. It is used as a measure of reproducibility or repeatability
of a segmentation method and denoted in Equation \ref{equ:DSC}. The average DSC is the mean of DSCs of FC, TC, FB and TB.
\begin{equation}
\begin{aligned}
\label{equ:DSC}
DSC(K,Y) = 100* (\frac{K\cap Y}{{\lvert K \rvert  +\lvert Y \rvert}})
\end{aligned}
\end{equation}

\noindent The VOE (\%) is commonly used for the evaluation of segmentation accuracy. A value of 0 denotes a perfect segmentation and a value of 100 means the segmentation and ground truth do not overlap as denoted in Equation \ref{equ:VOE}. The average VOE is the mean of VOEs of FC, TC, FB and TB.
\begin{equation}
\begin{aligned}
\label{equ:VOE}
VOE(K,Y) = 100* (1 - \frac{DSC}{200-DSC})  
\end{aligned}
\end{equation}

\noindent {The HD (mm) is a metric used to measure the similarity between two surfaces by measuring the maximum of minimum euclidean distance from a border voxel in surface K to a nearest voxel in surface Y and vise-versa as denoted in Equation} \ref{equ:HD}. {The value of 0 represents perfect segmentation while the larger value indicates greater dissimilarity between two sets.}

\begin{equation}
\begin{aligned}
\label{equ:HD}
H(K, Y) = \max_{k \in K} \min_{y \in Y} d(k, y)
\end{aligned}
\end{equation}

{Where, k and y are the points from set K (Predicted mask) and set Y (Ground truth mask) respectively and d(k,y) denotes euclidean distance between k and y.}\\

\noindent The PA (\%) is the percent of correctly classified pixels to all predicted pixels in the image as denoted in Equation \ref{equ:PA}. For binary mask; a true positive (TP) represents a pixel that is correctly predicted to belong to the given class (according to the target mask) whereas a true negative (TN) represents a pixel that is correctly identified as not belonging to the given class. False positive (FP) and false negative (FN) represents the wrongly classified pixels.
\begin{equation}
\begin{aligned}
\label{equ:PA}
PA = 100* (\frac{TP + TN}{TP + TN + FP + FN}) 
\end{aligned}
\end{equation}

\noindent{\textit{Qualitative metrics}}\\
The segmentation results are qualitatively compared using the boxplot and segmentation error map. The boxplot plots describe the skewness, dispersion, and outliers within the predicted data samples for DSC. The segmentation error
map is a pictorial representation of pixel-wise segmentation inaccuracies in the predicted map
compared with ground truth.\\

\noindent{\textit{Statistical Analysis metrics}}

\noindent The segmentation results are statistically validated using measures of association and measures of concordance. The intra-class correlation coefficient (ICC) 
and Pearson's correlation coefficient (r) 
are used as a measure of association to quantify the degree of relationship between automated segmentation and manual segmentation. While Kendall's coefficient of concordance (W) 
is used as a measure of concordance to quantify the level of agreement between automated segmentation and manual segmentation.

\subsection{Experimental Setup}
The data size of 273 subjects (32,760 slices) is utilized for training and validation and 108 subjects (17,280 slices) for model testing. Each MRI slice is assigned a SLICE ID before testing, corresponding to the subject ID (the first six digits) given by the OAI consortium, followed by the corresponding slice number. The proposed MtRA-Unet model is trained in an end-to-end manner, without any post-processing technique and by removing the initial and ending 20 slices. 
A simple thresholding technique is adopted for the ROI calculation and segmentation evaluation. According to this criterion, the threshold for minimum pixel count is kept at 300 and 100 for the FB, TB and FC, TC respectively. The model is trained for 100 epochs on MRI slice size of 150*150 and batch size of 150. The learning rate of 0.03 and adam optimizer is adopted to avoid overfitting. The model is trained and tested on an NVIDIA A100 80 GB GPU. Following testing, the model's optimal values [0.1,0.2,0.3,0.4] for $\lambda$, [0.01,0.1,0.27,0.12,0.5] for $\alpha$, 5 for m and 4 for n are utilized for the MtRA-Unet model. The detailed network configuration is reported in table \ref{tab:T3}.
For binary segmentation, the proposed model is trained for 100 epochs on an MRI slice size of 150*150 and batch size of 64. Following testing, the model's optimal values of 1 for $\alpha$, [0.1,0.2,0.3,0.4] for $\lambda$, 1 for m and 4 for n are utilized for the MtRA-Unet model. By Deng et al. \cite{deng2021coarse} criterion, the minimum pixel threshold is kept at 280 and 100 for FC and TC respectively. The segmentation accuracy is measured using DSC, VOE, HD, and PA for each sample as described in the evaluation metrics section.

\begin{table}
\caption{Parameters of proposed MtRA-Unet architecture}
\label{tab:T3}
\begin{tabular}{lc}
\toprule
Parameter & Value\\
\midrule
 Input shape & 150*150 \\
 Target spacing (mm) & 0.7*0.36*0.36 \\
 Image size & 384*384 \\
 Training Epochs & 100\\
 Learning Rate & $1*10^{-3}$\\
 Batch Size & 150\\
 Train/Validation/Test & 218/55/108\\
\botrule 
\end{tabular}
\end{table}

\section{Results}
To assess the outcomes of the suggested approach, both quantitative, qualitative, and statistical analyses are conducted.

\subsection{Quantitative analysis}
\begin{figure*}
    \centering
    \includegraphics[scale=0.30]{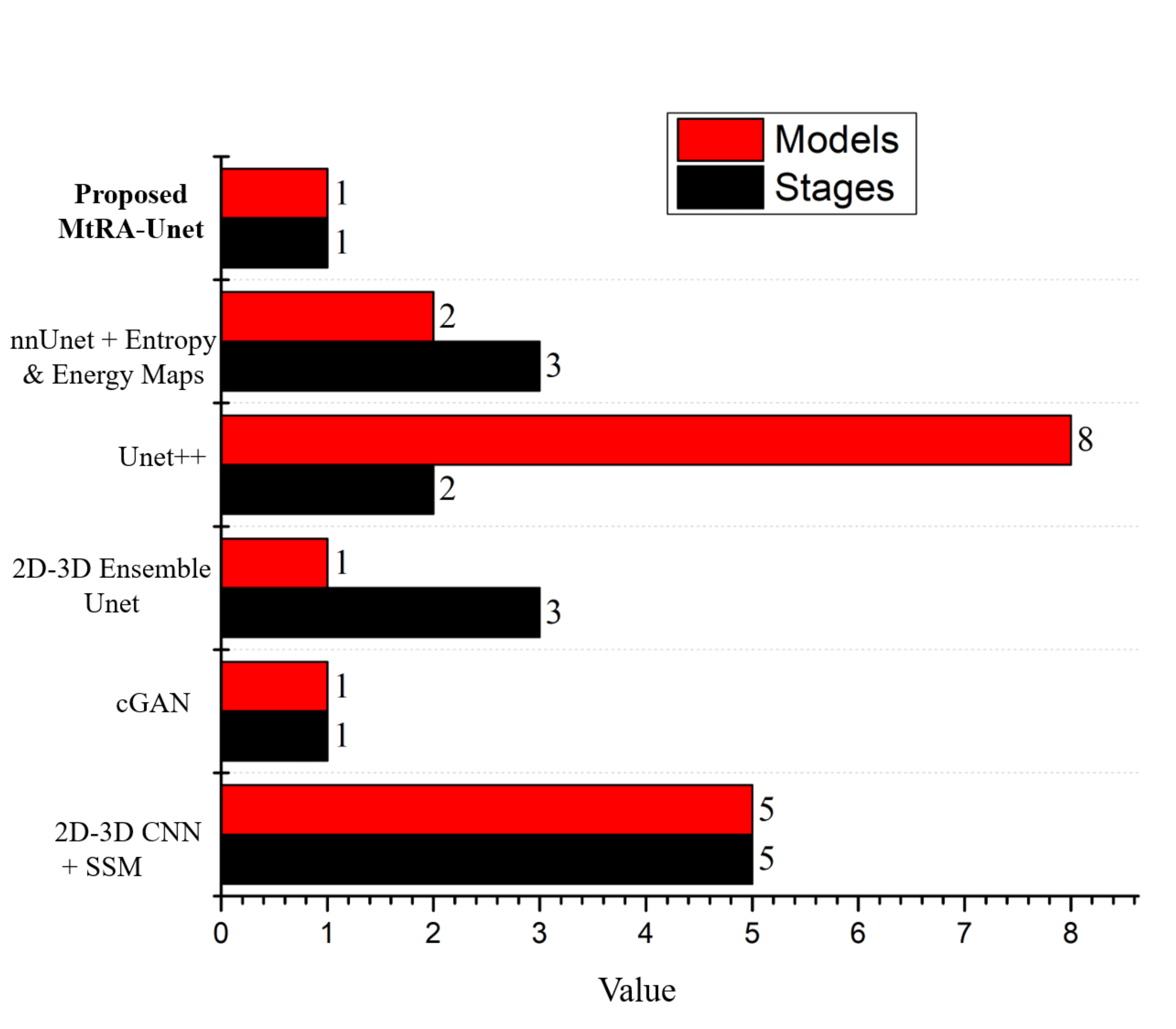}
    \caption{Comparison of the proposed work with state-of-the-art for number of parameters and a number of stages. Overall, the proposed work requires the least amount of processing} 
    \label{fig:Numpar}
\end{figure*}

\noindent\textit{Comparison with the state-of-the-arts}\\
The proposed work is compared with the recently published literature on knee tissue segmentation \cite{ambellan2019automated},\cite{abd2021automated},\cite{li2022entropy},\cite{deng2021coarse},\cite{kessler2020optimisation} involving the OAI-ZIB dataset. The quantitative comparison is separated into two parts; (1) a comparison of the segmentation pipeline and (2) a comparison of segmentation performance. 
The comparison of the segmentation pipeline includes a comparison of the total number of model parameters, preprocessing and postprocessing operations, the requirement of manual interference, segmentation time per subject and the number of models and stages as reported in Table \ref{tab:parameters} and Figure \ref{fig:Numpar} respectively.   
The manual interference includes the requirement of monitoring a multi-stage-segmentation pipeline in a fine-tuning stage such as for SSM-based approach \cite{ambellan2019automated}, entropy maps-based approach \cite{li2022entropy} and removing noisy data \cite{abd2021automated}, \cite{deng2021coarse}. The total model parameters in the segmentation pipeline, expressed in millions (M), were computed by implementing the provided architecture in Python. The total segmentation time/subject indicates the amount of time required to segment a single subject data volume (i.e 160 slices). It is noted that segmentation time/subject is reflected in Table \ref{tab:parameters} based on data mentioned in the corresponding publication. However, some of the architectures \cite{abd2021automated},\cite{deng2021coarse}, \cite{li2022entropy} may require additional data handling time apart from the mentioned segmentation time/subject.
The number of models greater than one represents the requirement of multiple models for binary segmentation of multiple tissues \cite{ambellan2019automated},\cite{deng2021coarse},\cite{li2022entropy} while the number of stages greater than one represents the total number of stages in segmentation pipeline that include coarse segmentation stage followed by segmentation refinement stage \cite{abd2021automated},\cite{deng2021coarse} and separate cartilage segmentation stage \cite{li2022entropy}. It is observed from the results that the highest preprocessing, postprocessing and manual interference is required by the architecture of Ambellen et al. \cite{ambellan2019automated} while the least is required for Kessler et al. \cite{kessler2020optimisation}, proposed MtRA-Unet and Latif at al. \cite{abd2021automated}. However, the architecture by Kessler et al. \cite{kessler2020optimisation} is highly dependent on the image quality during image generation. While the architecture by Latif et al. \cite{abd2021automated} requires computationally higher preprocessing operations. The proposed MtRA-Unet architecture was found to have a minimum number of models and stages (i.e. 1) and requires no postprocessing and manual interference. In addition, the proposed MtRA-Unet architecture is found to consume only 22 sec segmentation time per subject thus making it the fastest end-to-end segmentation pipeline for knee tissue segmentation.  As a result, even though the total number of parameters (i.e. 96.5 M) is on the higher side, the proposed scheme is highly suitable for clinical settings.

\begin{sidewaystable}

\caption{Comparison of segmentation pipeline of the proposed work with State-of-arts}\label{tab:parameters}
\begin{tabular*}{\textheight}{@{\extracolsep\fill}lcllcc}
\toprule%
Architecture & N & PR & PO & MI & Time/Subject\\&&&&& (sec)\\
\midrule
 2D and 3D CNN + SSM \cite{ambellan2019automated} & 79 M & 1. Ground truth binary & 1. SSM based $\blacktriangle$ $\blacktriangle$& \textcolor{black}{\checkmark} & 562 \\ && mask generation $\blacktriangle$ \\
 
 cGAN \cite{kessler2020optimisation} & 78* M & 1. Conversion of MR images & {1. Thresholding} {$\blacktriangledown$} & \textcolor{black}{\checkmark} & \textcolor{red}{{21\#}}\\ & \& 279* M & and masks in dicom (.dcm),\\ && Nifti files to portable\\&& network graphics (.png) files $\blacktriangledown$\\&& 2. Removing Noisy data $\textcolor{black}{\bullet}$\\
 
 2D-3D Ensemble Unet \cite{abd2021automated} & 134 M & 1. Mask reorientation $\blacktriangle$ & Not required & X & 62\#\# \\ && 2. Intensity normalization $\blacktriangle$\\ && 3. Resampling $\blacktriangle$ \\
 
 Unet++ \cite{deng2021coarse} & 73 M & 1. Ground truth binary & 1. Thresholding $\blacktriangledown$ & \textcolor{black}{\checkmark} & 36\#\#\\ && mask generation $\blacktriangle$ & 2. Binary mask \\ && 2. Image cropping & results ensembling $\blacktriangledown$\\ && before $2^{nd}$ stage $\blacktriangledown$\\
 
 nnUnet + Entropy \& Distance Maps \cite{li2022entropy} & ** & 1. MRI image cropping $\blacktriangledown$ & ** & \textcolor{black}{\checkmark}& 33.6\#\#\#\\ && 2. Ensembling entropy, distance \\&& maps before $3^{rd}$ stage $\blacktriangle$\\ && 3. Concatenated data cropping $\blacktriangle$\\
 
 \textbf{Proposed MtRA-Unet} & 96.5 M & 1. Conversion of .dcm, & Not required & X & \textcolor{blue}{22}\\ && metaImage metaHeader(.mhd) \\&& files to .png files $\blacktriangledown$\\ && 2. MR image and mask renaming $\blacktriangledown$\\
 
\botrule
\end{tabular*}
\footnotetext{Note: \item N - Number of model parameters, PR - Preprocessing operations, PO - Postprocessing Operations, MI - Requirement of Manual Interference}
\footnotetext{* Min. channels 64 \& 128, ** Not mentioned in the paper and neither code is available online, \# Calculated based on " $\sim$ 0.13 sec" segmentation time consumed per slice on the local dataset (Advanced MRI of Osteoarthritis, AMROA), \#\# Calculated based on 12 sec time consumed per tissue segmentation, \#\#\# Calculated based on 16.8 sec time consumed for FC and TC segmentation}
\footnotetext{$\blacktriangle$ Moderate computation, $\blacktriangledown$ Low computation, $\textcolor{black}{\bullet}$ Computation with manual input, $\blacktriangle$ $\blacktriangle$ High computation \item {$\divideontimes$ The best and second best results are denoted in} \textcolor{red}{{red}} {and} \textcolor{blue}{{blue}} {colors, respectively.}}
\end{sidewaystable}

\begin{table}[h!]
\caption{{Comparison of results of the proposed work with the State-of-the-art using Dice Similarity Coefficient (DSC), Volume Overlap Error (VOE), and Hausdorff Distance (HD) for bone and cartilage segmentation.}}\label{tab:T6}
\caption{Comparison of segmentation pipeline of the proposed work with State-of-arts}\label{tab:compare}
\begin{tabular*}{\textwidth}{@{\extracolsep\fill}lccccc}
\toprule%
Architecture & Metrics & FC & TC & FB & TB\\
\midrule
 2D and 3D CNN + SSM \cite{ambellan2019automated}& DSC (\%) $\uparrow$ & 89.9 & 85.6 & 98.5 & 98.5\\  & VOE (\%) $\downarrow$ &\textcolor{blue}{18.1}& 24.9 & \textcolor{red}{2.8} & \textcolor{blue}{2.9} \\ & HD (mm) $\downarrow$ & \textcolor{blue}{{5.35 }} & {6.35 } & \textcolor{red}{{2.93 }} & \textcolor{red}{{ 3.16 }}\\ 
 
cGAN \cite{kessler2020optimisation} & DSC (\%) & 89.5 & 83.9 & 98.5 & 98.5 \\ & VOE (\%) & \textcolor{blue}{18.92}& 27.55 & -- & --\\
 
2D-3D ensemble Unet \cite{abd2021automated} & DSC (\%) & \textcolor{blue}{90.3}& \textcolor{red}{86.5} & \textcolor{blue}{98.6} & \textcolor{red}{98.8}\\& VOE (\%) & \textcolor{red}{17.5} & \textcolor{red}{23.6} & \textcolor{red}{2.8} & \textcolor{red}{2.4}\\ 
 
Unet++ \cite{deng2021coarse} & DSC (\%) & \textcolor{red}{90.9} & 85.8 & \textcolor{red}{99.1} & 98.2\\
 
nnUnet + Entropy \& Distance Maps \cite{li2022entropy} & DSC (\%)& 89.8  & \textcolor{blue}{86.4} & \textcolor{blue}{98.6} & \textcolor{blue}{98.6} \\ & {HD (mm)} & \textcolor{red}{{5.22 }} & \textcolor{blue}{{4.70 }} & {11.82 } & { 5.30 }\\

\textbf{Proposed MtRA-Unet} & DSC (\%) & 89.1 & 86.1 & 98.5 & 98.4 \\ & VOE (\%)& 19.18 &\textcolor{blue}{23.70}&\textcolor{blue}{2.92}& 3.09 \\ & {HD (mm)} & {5.49} &  \textcolor{red}{{4.04}} &  \textcolor{blue}{{4.43}} &  \textcolor{blue}{{3.23}}\\ 
\botrule
\end{tabular*}
\footnotetext{ {$\divideontimes$ The best and second best results are denoted in} \textcolor{red}{{red}} {and} \textcolor{blue}{{blue}} {colors, respectively.}}
\end{table}

\begin{figure*}
    \centering
    \includegraphics[scale=0.45]{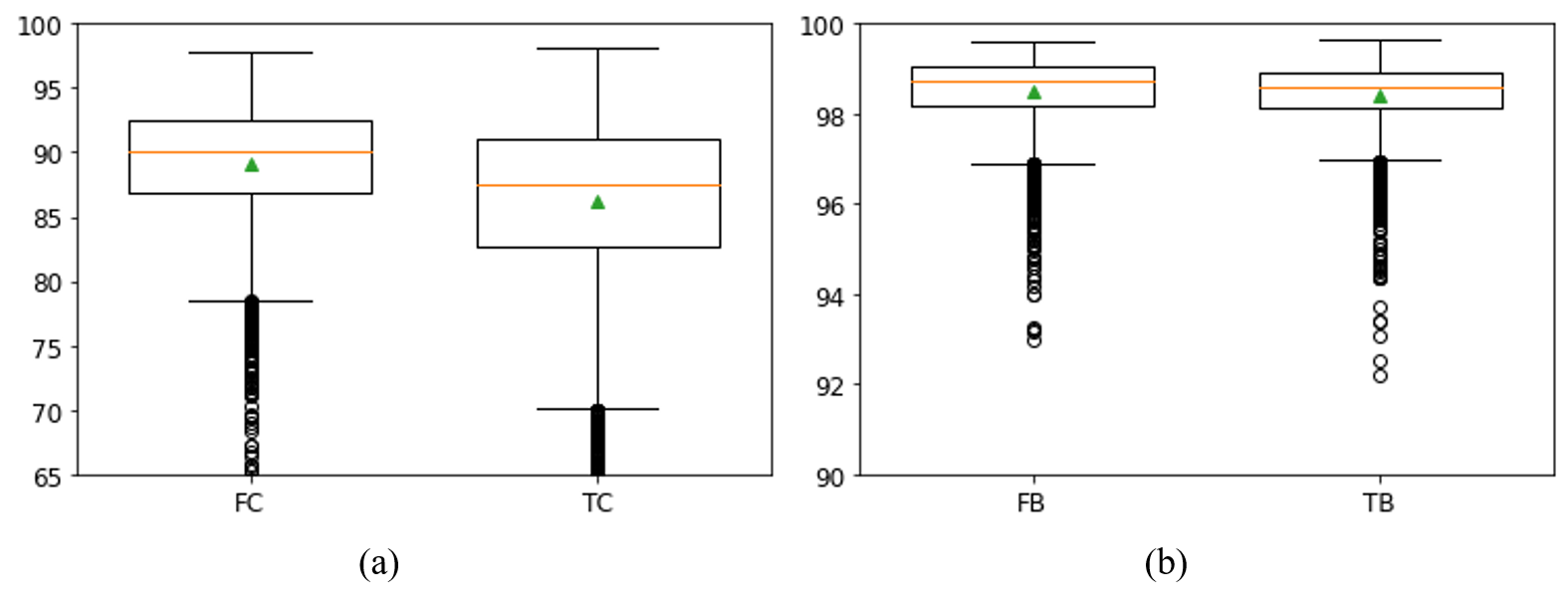}
    \caption{{Boxplot representation of DSC distribution per sample for (a) Femoral Cartilage (FC) and Tibial Cartilage (TC) and (b) Femoral Bone (FB) and Tibial Bone (TB). Overall, all boxes are ranged over a narrow region}} 
    \label{fig:QAB}
\end{figure*}

The comparison of segmentation performance is reported in Table \ref{tab:compare}. It indicates that the proposed MtRA-Unet model has produced results that are quite close to the state-of-art in terms of segmentation metrics for FB, TB and TC except for the FC. The TC results are nearly 1\% DSC higher than Ambellen et al. \cite{ambellan2019automated}, Kessler et al. \cite{kessler2020optimisation} and Deng et al. \cite{deng2021coarse} while for FB and TB the results are quite same as others. {Also, the HD of the TC is found to be 14\% superior as compared to Li et al.} \cite{li2022entropy} {which was specifically designed for small tissue segmentation.} However, the lower results for FC are also observed which might be because of imbalanced context learning using MRFF and attention module CBAM. The 50\% distribution of DSC for FC, TC, FB and TB per subject is found to accumulate in the narrow region in the boxplot with a maximum inter-quartile range of 8\% as depicted in Figure \ref{fig:QAB}. Hence, the predicted data was found to be less dispersed and symmetrically distributed for all the tissues that support the congruence of the proposed work with the reference standards.\\

\noindent\textit{Comparison with baseline network architectures}\\
In this experiment, the MtRA model is compared with several baseline segmentation networks
such as Segnet \cite{badrinarayanan2017segnet}, High-resolution network \cite{wang2020deep} and Bisenet
\cite{yu2018bisenet} networks as reported in Table \ref{tab:T4}. The comparison is performed by considering the weighted dice loss \cite{sorensen1948method} with a similar hyperparameter setting and computational resources as mentioned in Table \ref{tab:T3}. The selection of Unet as baseline architecture is justified with less than a half segmentation time/subject (2.9 sec) and higher pixel accuracy (98.89\%) as compared to other baseline architectures, which is critical in medical diagnosis. 
{Utilizing the Unpaired t-test at a significance level of 0.05, the analysis revealed highly significant statistical improvement (p\_value $<$ 0.001) in the performance of the proposed MtRA-Unet model compared to other segmentation models for FB, TB, and FC in terms of DSC and FB in terms of HD.}

\begin{table}[h!]
\caption{{Comparison of the proposed MtRA-Unet model results with other network architectures by Dice similarity coefficient (DSC), Volume overlap Error (VOE), and Hausdorff Distance (HD). ‘*” (p $<$ 0.001) and "**" (p $<$ 0.05) indicate statistically significant improvement by the proposed "MtRA-Unet" model, compared with other methods with a significance level of 0.05}}\label{tab:T4}

\begin{tabular*}{\textwidth}{@{\extracolsep\fill}lcccccc}
\toprule%
Architecture &Metrics&FC & TC & FB & TB & PA \\&&&&&& (\%)\\
\midrule
 Segnet \cite{badrinarayanan2017segnet} & DSC (\%) & 87.5 & 85.3 & 97.8 & 97.8 & 98.63 \\&  VOE (\%)  & 21.79 & 24.94 & 4.17 & 4.10 &\\& HD (mm)& {8.99} & {4.96} & {17.50} & {11.77} & \\ 

 BiseNet \cite{yu2018bisenet} & DSC (\%)  & 87.3 & 85.3 & 97.2 & 97.7 & 98.27 \\& VOE (\%) & 31.25 & 27.04 & 5.12 & 5.04 &\\& HD (mm) & {9.12} & {5.11} & {18.5} & {11.72} & \\

 HRnet \cite{wang2020deep} & DSC (\%) & \textcolor{blue}{88.4} & \textcolor{red}{86.5} & \textcolor{blue}{98.2} & \textcolor{blue}{98.2} & 
 98.88\\& VOE (\%) & \textcolor{blue}{20.40} & \textcolor{red}{23.21} & \textcolor{blue}{3.33} & 3.38 & \\ & HD (mm)& \textcolor{red}{{5.28}} & \textcolor{red}{{3.94}} & \textcolor{blue}{{6.90}} & \textcolor{red}{{3.64}} & \\  

 Unet \cite{ronneberger2015u} & DSC (\%) & 88.2 & \textcolor{blue}{86.2} & \textcolor{blue}{98.2} & \textcolor{blue}{98.2} & \textcolor{blue}{98.89} \\ & VOE (\%) & 20.61 & \textcolor{blue}{23.67} & \textcolor{blue}{3.33} & \textcolor{red}{3.33} &\\& HD (mm) & {6.54} & {4.50} & {16.10} & {5.46} & \\ 

 \textbf{Proposed} & DSC (\%) & \textcolor{red}{88.5}{*} & 85.8 & \textcolor{red}{98.3}{*} & \textcolor{red}{98.3}{*} & \textcolor{red}{98.95}{*} \\ \textbf{MtRA-Unet} & VOE (\%) & \textcolor{red}{20.13}{**} & 26.34 & \textcolor{red}{3.17} & \textcolor{blue}{3.36} &\\& HD (mm) & \textcolor{blue}{5.40}& \textcolor{blue}{{4.09}}& \textcolor{red}{{6.11}}{*} & \textcolor{blue}{{4.22}}& \\ 
\botrule
\end{tabular*}
\footnotetext{ $\divideontimes$ The best and second best results are highlighted with} \textcolor{red}{{red}} {and} \textcolor{blue}{{blue}} {colors, respectively.}
\end{table}

\subsection{Qualitative analysis}
\begin{figure*}[h!]
    \centering
    \includegraphics[scale=0.3]{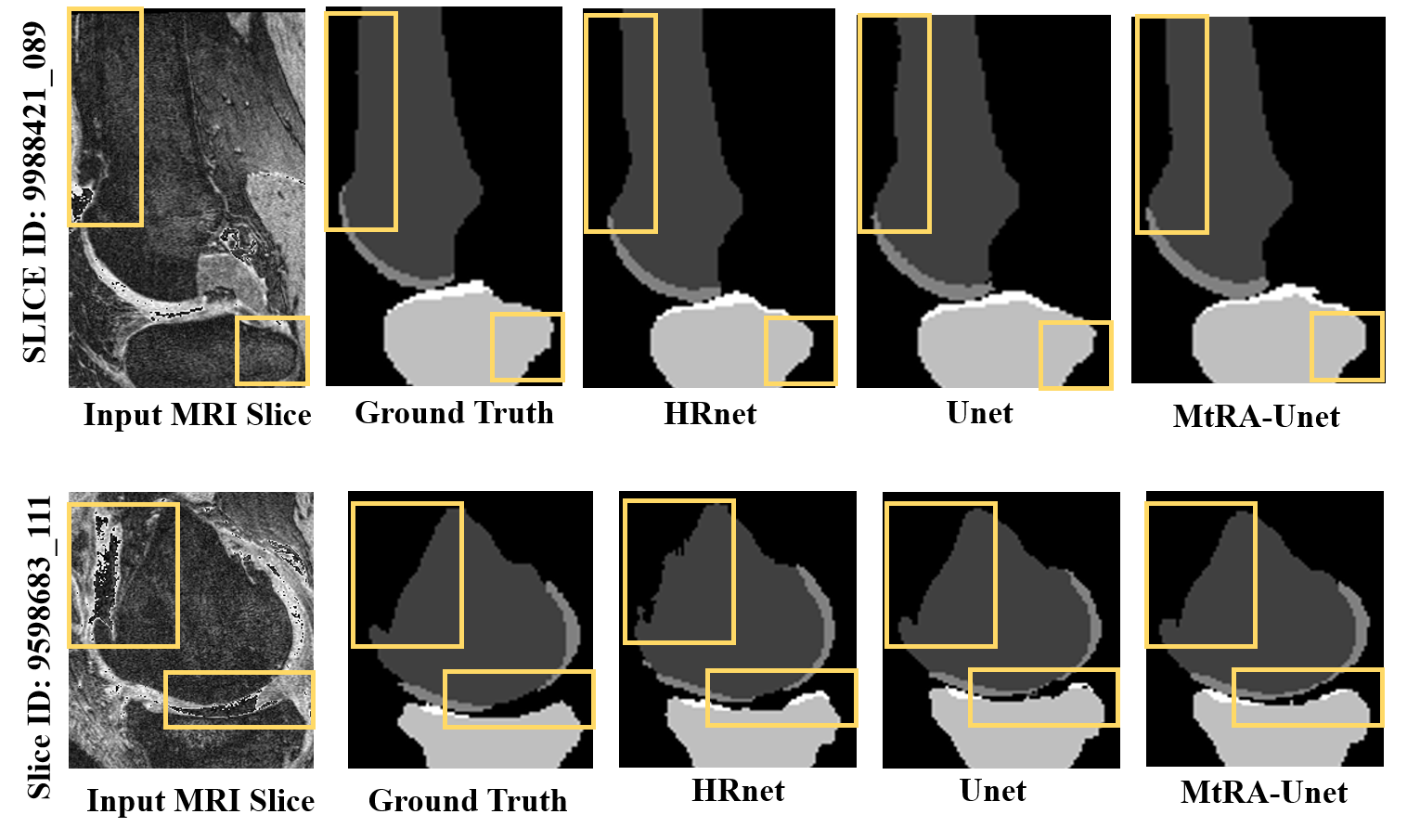}
    \caption{Visualization of qualitative results of HRnet, Unet and the proposed MtRA-Unet model for SLICE ID: 9988421\_089 and 9598683\_111} 
    \label{fig:QA3}
\end{figure*}

The experimentation results are supported by the qualitative analysis as reported in Figure \ref{fig:QA3}. The yellow bounding box on the MRI slice and segmentation map highlights those areas of MRI where visual agreements or disagreements are perceived. The MtRA-Unet model is found to restore the structural details of the femur and tibia as compared to HRnet and Unet as shown in Figure \ref{fig:QA3}. Also, no noticeable improvements are observed in the structure of cartilages despite the quantitative difference in the results of Unet and MtRA-Unet.


\subsection{{Statistical Analysis}}

{The measures of agreement between manual and automatic segmentations are measured using the F-statistic of one-way Analysis of Variance (ANOVA) with 0.05 significance level as denoted in Table} \ref{tab:MA}. {As seen from the statistical analysis, the FB and TB are highly correlated and reliable (p\_value $<$ 0.001) while the FC and TC correlated the least (p\_value $<$ 0.001) with moderate-to-high reliability. The less correlation and concordance of FC and TC may be due to its intricate shape, small size, and inconstant boundaries.}\\

\begin{table}[h!]
\caption{{Comparison of Measures of agreement between automatic segmentation and manual segmentation delineation. All measures; Pearson's Correlation Coefficient (r), Intra-class Correlation Coefficient (ICC) and Kendall's Coefficient of Concordance (W) are significant with p $<$ 0.001 and p** $<$ 0.05) }}
\label{tab:MA}
\begin{tabular*}{\textwidth}{@{\extracolsep\fill}lccccc}
\toprule%
Metrics & & FC & TC & FB & TB \\
\midrule
 Pearson's & \textbf{r} & 0.90 & 0.88{**} & 0.98 & 0.98\\
 McGraw's & \textbf{ICC} & 0.93 & 0.92 & 0.99 & 0.99\\
 Kendells's & \textbf{K} & 0.99 & 0.99 & 0.99 & 0.99\\
\botrule
\end{tabular*}
\end{table}

\section{Ablation Study}

To gauge the robustness of the proposed scheme, an ablation study is undertaken. This study explored the impact of varying loss function weights, compared different MRFF modules, assessed the effects of training sizes, and examined the binary segmentation of cartilages.\\

\subsection{Effect of different Loss Function Weights}
\begin{figure*}[h!]
    \centering
    \includegraphics[scale=0.25]{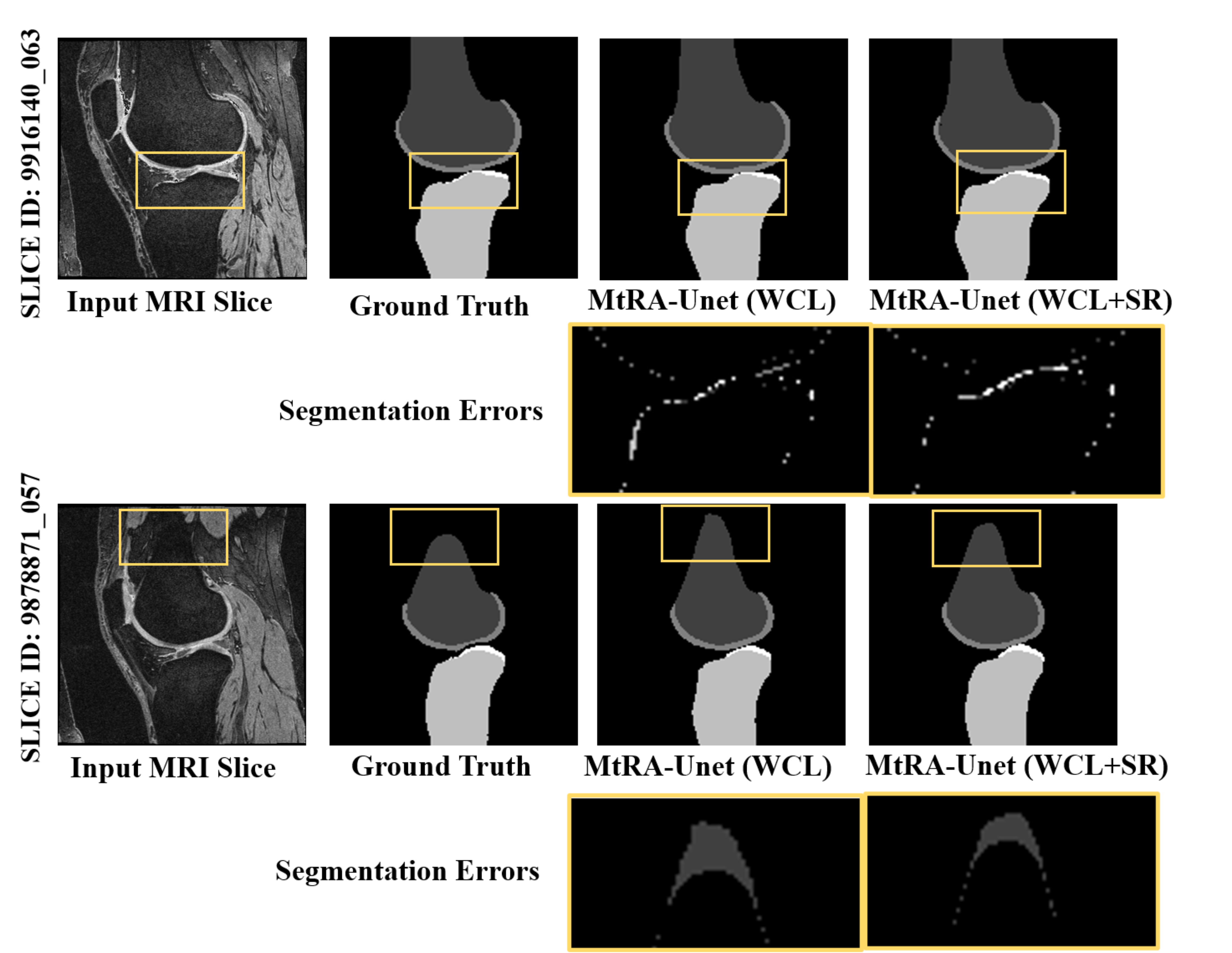}
    \caption{Visualization of qualitative results of the proposed MtRA-Unet model with and without shape reconstruction loss (SRL) function for femur and tibia (SLICE ID: 9916140\_063 and 9878871\_057)} 
    \label{fig:QA2}
\end{figure*}

The experiment is conducted on the MtRA-Unet model for a variety of loss function weights as indicated in Table \ref{tab:T9}. It is observed that the optimal settings for $\eta$ and $\gamma$  are 0.3 and 0.7 respectively. The significance of using SRL is confirmed by qualitative analysis with segmentation error maps of the tibia and femur which indicate lower segmentation error for a tibial and femoral region for MtRA-Unet with WCL+SRL as compared to the MtRA-Unet model with WCL as shown in Figure \ref{fig:QA2}. But, it is observed that the higher weight for SRL resulted in a poor model's performance with a decrease of 1\% of average DSC over the optimal network setting. 

\begin{table}[h!]
\caption{Comparison of the proposed MtRA-Unet model results with varying loss function weights by Average Dice Similarity Coefficient (DSC) and Pixel Accuracy (PA)}\label{tab:T9}
\begin{tabular*}{\textwidth}{@{\extracolsep\fill}lcccc}
\toprule
Model & $\eta$/$\gamma$ & Average DSC (\%) & Average VOE(\%) & Pixel Accuracy (\%)\\
\midrule
 1 & 0.1/0.9 & 92.9 & \textcolor{blue}{12.44} & 98.93\\
 2 & 0.3/0.7 & \textcolor{blue}{93.0} & \textcolor{red}{12.35} & \textcolor{blue}{99.00}\\
 3 & 0.5/0.5 & \textcolor{red}{93.2} & 13.01 & 98.98\\
 4 & 0.7/0.3 & \textcolor{blue}{93.0} & 12.75 & \textcolor{red}{99.02}\\
 5 & 0.9/0.1 & 92.2 & 12.98 & \textcolor{blue}{99.00}\\
\botrule
\end{tabular*}
\footnotetext{ $\divideontimes$ The best and second best results are highlighted with} \textcolor{red}{{red}} {and} \textcolor{blue}{{blue}} {colors, respectively.}
\end{table}

\subsection{Comparison with different MRFF modules}
The experiment is conducted on a variety of MRFF modules as indicated in Table \ref{tab:T5}. The evaluation is performed for WDL, WCL and the combined loss function with optimal $\gamma$ and $\eta$ values. The MRFF module contains the concatenated feature map that is multiplied by the feature map obtained after the GAP module. This multiplied feature map is then added to the original feature map. In MRFF1, the GAP module is omitted, while in MRFF2, it is applied in series with the concatenated feature map. It is observed that the combined loss function resulted in a better performance than the pixel-wise loss functions for all MRFF combinations with pixel accuracy $>=0.5$\%. The superior results are achieved from the MtRA-Unet model with PA of 99.00\%, DSC of bones with 98.5\% {(p\_value $<$ 0.001)} for FB, and 98.5\% {(p\_value $<$ 0.001)} for TB. The improved DSC results for MRFF2 and MtRA-Unet over MRFF1 have motivated the selection of the GAP module and its placement in parallel to the concatenated feature map.\\

\begin{table}[h!]
\caption{{Comparison of proposed MtRA-Unet model results for different Multi-Resolution Feature Fusion (MRFF) modules by Dice Similarity Coefficient (DSC), Volumetric Overlap Error (VOE), and Housdorff Distance  (HD). ‘*” (p $<$ 0.001) and "**" (p $<$ 0.05) indicate statistically significant improvement by the proposed "MtRA-Unet" model, compared with other methods with a significance level of 0.05}}
\label{tab:T5}
\begin{tabular*}{\textwidth}{@{\extracolsep\fill}lccccccc}
\toprule%
Architecture &Loss &Metrics &FC & TC & FB & TB & PA \\& Function &&&&&&(\%)\\
\midrule
  $\text{baseline}^{ \dagger}$+ MRFF1 & WDL & DSC (\%) & 88.8 & 86.1 & \textcolor{blue}{98.4} & \textcolor{blue}{98.3} & 98.93 \\ && VOE (\%) &19.73&\textcolor{red}{22.60}&3.19&3.18&\\& & {HD (mm)} & {7.31} & {4.33} & {12.80} & {4.86} & \\
  
  & WCL & DSC (\%) & 88.6 & 86.0 & \textcolor{blue}{98.4} & \textcolor{blue}{98.3} & 98.94 \\ &&  VOE (\%) & 19.94 & 23.83 & 3.11 & 3.31 &\\&& {HD (mm)} & \textcolor{red}{{5.24}} & {4.75} & {5.62}&{4.26}& \\
  
  & WCL+SRL & DSC (\%) & 88.9 & \textcolor{blue}{86.6} & \textcolor{red}{98.5} & \textcolor{red}{98.4} & 98.97 \\ && VOE (\%) & 19.54 & 22.97 & 3.00 &\textcolor{red}{3.08}&\\&& {HD (mm)} & {6.07} & \textcolor{blue}{{4.07}} & \textcolor{blue}{{4.60}} & \textcolor{blue}{{3.15}} &\\
 
  $\text{baseline}^{ \dagger}$+ MRFF2 & WDL & DSC (\%)& \textcolor{blue}{89.0} & \textcolor{red}{86.7} & 98.3 & \textcolor{blue}{98.3} & 98.91\\ && VOE (\%)&19.44&\textcolor{blue}{22.73}&3.21&3.31&\\&& {HD (mm)}  & {6.03} & {4.23} & {5.05} & {3.25}& \\
 
 & WCL & DSC (\%) & 88.9 & 86.5 & 98.2 & \textcolor{blue}{98.3} & 98.92 \\ &&  VOE (\%) & 19.42 & 23.03 & 3.33 & 3.16 &\\&& {HD (mm)}  & {11.17} & {7.40} & {5.63}&{4.47}& \\
 
 & WCL+SRL & DSC (\%) &\ 88.9 & 86.2 & \textcolor{blue}{98.4} & \textcolor{blue}{98.3} & \textcolor{red}{99.04} \\ && VOE (\%) &19.47&23.50&\textcolor{blue}{2.99}&3.18& \\&& {HD (mm)} & {7.62} & {4.47} & {6.25}&{3.51}&\\ 
 
 $\text{baseline}^{ \dagger}$ + MRFF & WDL & DSC (\%)
 & 88.5 & 85.8 & \textcolor{blue}{98.3} & 98.3 & 98.95 \\(\textbf{Proposed MtRA-Unet}&&  VOE (\%) & 20.13 & 26.34 & 3.17 & 3.36 &\\ && {HD (mm)} & \textcolor{blue}{{5.40}} & \textcolor{blue}{{4.09}} & {6.11} & {4.22} & \\
 
 & WCL & DSC (\%) & \textcolor{red}{89.1} & \textcolor{blue}{86.6} & \textcolor{blue}{98.3} & \textcolor{blue}{98.3} & 98.93\\ &&VOE (\%) &\textcolor{blue}{19.26}&22.99&3.21&3.17&\\&& {HD (mm)}  & {6.41} & {5.83} & {6.94} & {4.37}& \\
 
 & WCL+SRL & DSC (\%) & \textcolor{red}{89.1}{*} & 86.1& \textcolor{red}{98.5}{*} & \textcolor{red}{98.4}{*} & \textcolor{blue}{99.00}\\ &&VOE (\%) &\textcolor{red}{19.18}{**}&23.70&\textcolor{red}{2.92}&\textcolor{blue}{3.09}&\\ && {HD (mm)}  & {5.49} & \textcolor{red}{{4.04}}{**} & \textcolor{red}{{4.43}}{*} & \textcolor{blue}{{3.23}} & \\
 \botrule
\end{tabular*}
\footnotetext{ $\divideontimes$ The best and second best results are highlighted with} \textcolor{red}{{red}} {and} \textcolor{blue}{{blue}} {colors, respectively.}
\end{table}

\subsection{Effect of training size}
In some of the MRI volumes in the OAI repository, the desired sequence is found to be absent or has failed to appear due to technical difficulties. Nevertheless, an ablation study is conducted for different training data sizes, as indicated in Table \ref{tab:Data_size}, to justify the insignificant performance changes. The series1 and series2 models are trained for different sampling strategies and an average of their results is indicated in Table \ref{tab:Data_size} while series 3 models are trained for 273 data volumes. It can be observed from the results that a minor segmentation performance change $<$ 1\% in DSC was obtained for FB and TB. However, the DSC results of FC and TC showed a significant change of $\sim$ 3\% due to data size variation. But once the data size is changed above to 273 for training there is a very minor change of $<$ 1.5\% in DSC as observed from previous literature \cite{kessler2020optimisation}, \cite{li2022entropy}, thereby indicating sufficiency of the mentioned training data size.\\

\begin{table}[h!]
\caption{Comparison of "MtRA-Unet" results for different training data sizes using Dice Similarity Coefficient (DSC), Volume Overlap Error (VOE), and Hausdorff Distance (HD)}
\label{tab:Data_size}
\begin{tabular*}{\textwidth}{@{\extracolsep\fill}lccccccc}
\toprule
Series Name & Train Data Size & Test Data Size & Metric (\%) & FB & TB & FC & TC\\
\midrule
1 & 150 & 108 & DSC (\%) & 98.0 & 97.8 & 86.4 &  83.1 \\&&& VOE (\%) &3.87&4.10&25.10&29.02\\ &&& HD (mm) &3.75&4.05&5.28&5.01\\
2 & 200 & 108 & DSC (\%) & 98.2 & 98.1  & 86.2 & 83.4\\&&& VOE (\%) &3.57&2.25&24.67&28.97\\&&& HD (mm) &5.10&3.68&7.18&5.86\\
3 & 273 & 108 & DSC (\%) & 98.5 & 98.4 & 89.1 & 86.1 \\&&& VOE (\%) &2.92 &3.09&19.18 &23.70 \\&&& HD (mm)  & 4.43 & 3.23 & 5.49 & 4.04\\
\botrule
\end{tabular*}
\end{table}

\subsection{Binary Segmentation of Cartilages}
The proposed model is verified on the binary segmentation of FC and TC with Unet and combined loss as shown in Table \ref{tab:FTC}. From the results of the FC segmentation, it is observed that the proposed MtRA-Unet model has achieved excellent performance with DSC of 89.3\% ({p\_value $<$ 0.001}, VOE of 19.02\% ({p\_value $<$ 0.001}, {HD of 3.72 mm} and PA of 99.30\% as reported in Table \ref{tab:FTC}. 

\begin{figure*}
    \centering
    \includegraphics[scale=0.55]{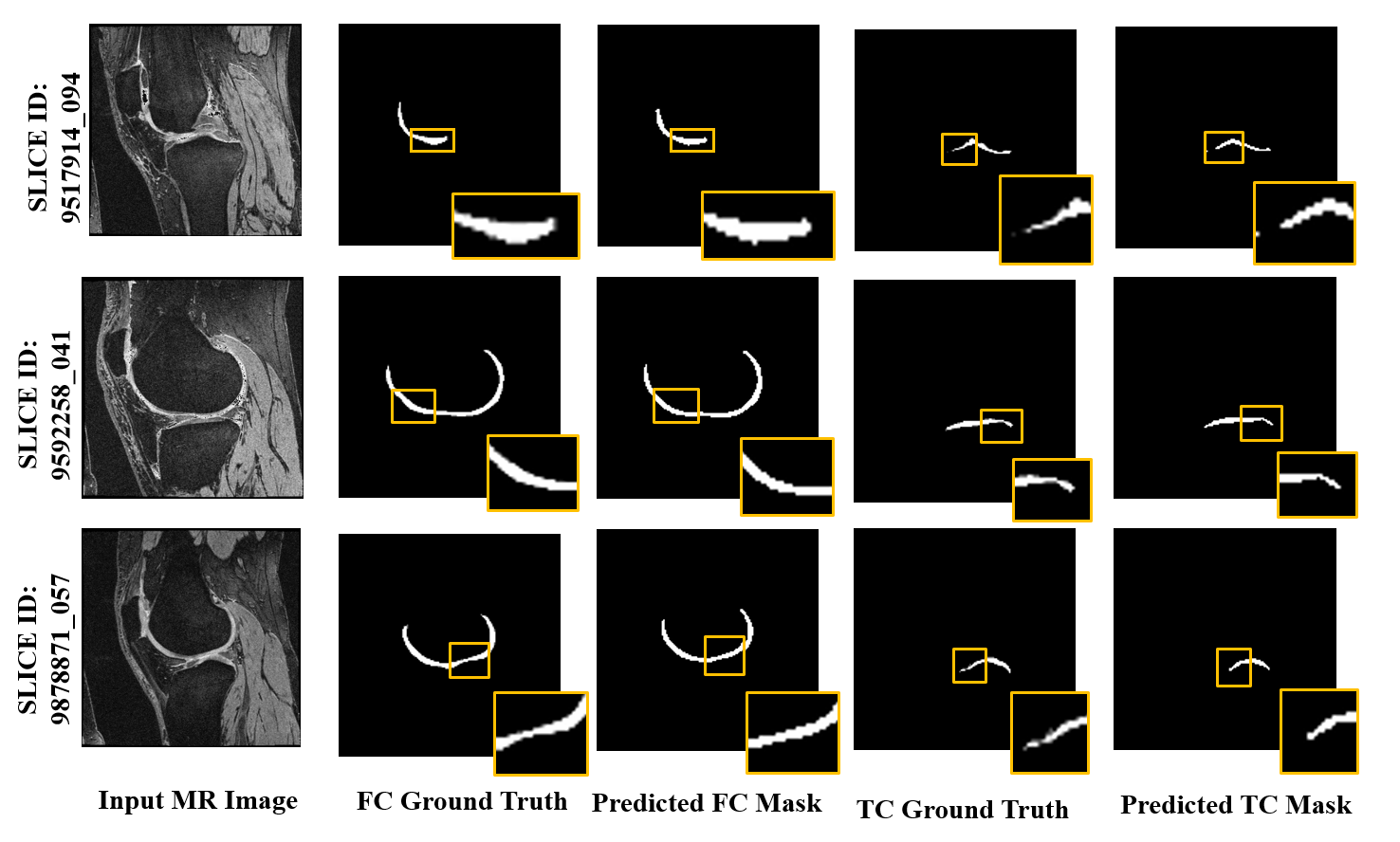}
    \caption{Visualization of qualitative analysis of the proposed work for Femoral Cartilage (FC) and Tibial Cartilage (TC) with enlarged ROI at the bottom right corner for SLICE ID: 9517914\_094, 9592258\_041 and 9878871\_057}
    \label{fig:QA_cart}
\end{figure*}

In the case of TC segmentation, the MtRA-Unet model is comparable with Unet by producing DSC of 86.4\%, and PA of 99.72\% with a combined loss function. No significant improvement is found in the segmentation of TC might be because of (1) lower shape variability, and (2) uniform thickness of tibial cartilage along the Bone-Cartilage Interface (BCI). In conclusion, the MtRA-Unet model with the combined loss function is observed to improve the boundary details, shape, and size details for critical regions of FC and TC as depicted in Figure \ref{fig:QA_cart}.\\

\begin{table}
\caption{{Comparison of the proposed MtRA-Unet model with Unet \cite{ronneberger2015u} (with and without Shape Reconstruction Loss (SRL)) for segmentation of Femoral Cartilage (FC) and Tibial Cartilage (TC) by Dice Similarity Coefficient (DSC), Volume Overlap Error (VOE), and Hausdorff Distance (HD). ‘*” (p $<$ 0.001) and "**" (p $<$ 0.05) indicate statistically significant improvement by the proposed "MtRA-Unet" model, compared with other methods with a significance level of 0.05}}
\label{tab:FTC}

\begin{tabular*}{\textwidth}{@{\extracolsep\fill}lcccc}
\toprule
Architecture & Loss & Metrics & FC & TC \\ & Function &&&\\
\midrule
Unet \cite{ronneberger2015u} & WCL & DSC (\%) & 89.0 & \textcolor{blue}{86.0} \\  && VOE (\%) & 19.52 & 23.90\\ && PA (\%) & \textcolor{blue}{99.29} & \textcolor{red}{99.72} \\ && {HD (mm)} & {16.67} & {17.64}\\

Unet \cite{ronneberger2015u} & WCL + SRL & DSC (\%) & \textcolor{blue}{89.1} & \textcolor{red}{86.4} \\&&  VOE (\%) & \textcolor{blue}{19.43} & \textcolor{red}{23.44} \\ && PA (\%)& \textcolor{blue}{99.29} & \textcolor{red}{99.72} \\ && {HD (mm)}  & \textcolor{red}{{3.54}} & \textcolor{red}{{2.58}}\\

Proposed MtRA-Unet & WCL+SRL & DSC (\%)& \textcolor{red}{89.3}{*} & \textcolor{red}{86.4} \\ && VOE (\%)& \textcolor{red}{19.02}{*} & \textcolor{blue}{23.47} \\ && PA (\%) & \textcolor{red}{99.30} & \textcolor{red}{99.72}{*} \\ && {HD (mm)} & \textcolor{blue}{{3.72}} & \textcolor{blue}{{2.78}}\\
\botrule
\end{tabular*}
\footnotetext{ $\divideontimes$ The best and second best results are highlighted with} \textcolor{red}{{red}} {and} \textcolor{blue}{{blue}} {colors, respectively.}
\end{table}

\section{Discussion}
The performance of the present work is compared with the state-of-the-art segmentation models on the OAI-ZIB dataset as shown in Table \ref{tab:compare}. Even though the proposed work is implemented with 381 MRI volumes still the results are comparable with most of the existing works including 2D CNN-based methods \cite{deng2021coarse}, 2D-3D CNN ensemble-based methods \cite{ambellan2019automated},\cite{abd2021automated}, generative \cite{kessler2020optimisation} methods while consuming only 22 sec segmentation time per MRI volume as depicted in Table \ref{tab:Data_size}. Thus reducing segmentation time/subject by 35\% to 96\% as compared to state-of-the-art. The existing methods are limited to handling large variations of shape and appearance and require frequent manual interference \cite{ebrahimkhani2022automated}. In addition, many of these methods use additional 3D features \cite{ambellan2019automated},\cite{abd2021automated} or cumbersome second-stage segmentation refinement techniques such as atlas-based statistical shape modeling \cite{ambellan2019automated}, diffeomorphic mapping \cite{ebrahimkhani2022automated} that consume a lot of time. Another major advantage of the proposed work is that the ROI slices per subject are identified for both medial and lateral regions that can provide dependable clinical assistance for KOA diagnosis to medical professionals.

The comparison and validation of the proposed MtRA-Unet model with different network architectures and different MRFF modules directed the selection of Unet as baseline architecture and utilization of GAP in MRFF module as reported in Table \ref{tab:T4} and \ref{tab:T5}. The multi-resolution feature convolution is used to capture the multi-scale contextual information of incoming features at multiple field-of views that address tissue-specific scale ambiguity issues and refine the spatial relationships between different tissue structures. The addition of CBAM was found to improve the results of FC and TC, which signifies the requirement of attention schemes for medical images that contain small tissue structures. However, for binary segmentation of tissues with lesser shape variability, such attention schemes were found to produce limited improvement in performance as described in Table \ref{tab:FTC}. By design, the dice loss \cite{sorensen1948method}
with cross-entropy loss empirically ameliorates the segmentation performance and model stability in medical images. However, in most of the implementations, the structural details are still left out due to non-delineating boundary details. In such cases, the unified pixel-wise and SR loss functions were found to reconfigure the structural details while segmentation of tissue structures as observed in Table \ref{tab:T5}. However, the improvement in overall segmentation performance is noticed when the proposed SR loss is assigned with a lower weight which indicates the need for fewer structural re-configurations while working with CNN as described in Table \ref{tab:T9}. Also, the use of SR loss can be perceived as the second stage segmentation refinement stage that needs no complex feature matching and registration steps \cite{ambellan2019automated} {which can produce uniform segmentation (lower HD) over the other loss functions as reported in Table} \ref{tab:T4} and \ref{tab:FTC}. The proposed work has achieved an overall DSC of 89.1\% for the FC,86.1\% for the TC and 98.5\% and 98.4\% for the FB and TB respectively. However, the robustness of the MtRA-Unet model is further confirmed by binary segmentation with DSC of 89.3\% for FC and 86.4\% for TC. 

Though our research shows vital implications in the analysis of KOA still it suffers from some drawbacks. One of the major drawbacks of the research is that the complete MRI data volume from OAI-ZIB is not utilized to compare the performance with state-of-the-art due to technical difficulties in the retrieval of desired MRI volumes. Also, the proposed work is not validated on the MRI data from any other data sources where high image quality may not be guaranteed. In addition, the thresholding strategy adopted in the presented work may not be suitable for data volumes differing in demographics and spatial resolution. Lastly, the presented work is not investigated for different hyper-parameters settings and network architectures. However, the possible extension would be to analyze the impact of severity grades and demographics on the segmentation performance with diverse data volumes.

\section{Conclusion}
In conclusion, the proposed work is an end-to-end single-stage segmentation network that learns hierarchies of multi-contextual features that focus on the most relevant features and learns inaccuracies between predictions and ground truth features directly from training data without needing any complex and time-consuming post-processing. In addition, the segmentation performance of the proposed work concentrates only on the ROI slices that can provide dependable assistance to medical professionals for clinical inference and disorder management. The time to segment a single MRI volume is reported to be 22 sec. which is fast enough for clinical application.

\section*{Declaration of competing interests}
The authors declare no competing interests.

\section*{Acknowledgement}
We would like to express our gratitude to the OAI and OAI-ZIB investigators for providing the dataset. We are also thankful to the Department of Biotechnology (DBT), Government of India (GOI), and  NorthEast Centre for Biological Sciences and Healthcare Engineering (NECBH) for providing high-end computational facilities ( Project Number - BT/COE/34/SP28408/2018). The manuscript does not represent the ideas or beliefs of the OAI investigators, the National Institutes of Health, the Zuse Institute Berlin, DBT-GOI and NECBH, IIT Guwahati.\\ 

\section*{CRediT authorship contribution statement}
\textbf{Akshay Daydar} Conceptualization, Methodology, Software, Investigation, Data curation, Visualization, Writing - original draft. \textbf{Alik Pramanick} Conceptualization, Data curation, Writing - review \& editing. \textbf{Arijit Sur} Validation, Writing - Review \& Editing, Supervision. \textbf{Subramani Kanagaraj} Validation, Writing - Review \& Editing, Supervision.
\bibliography{sn-article}

\end{document}